\magnification=1200
{\nopagenumbers
\baselineskip=15pt
\hsize=5.5in
\hskip10cm UM--P-97/26\hfil\par
\hskip10cm RCHEP--97/04 \hfil
\vskip2cm
\centerline{\bf The Mass Spectrum of Light and Heavy Hadrons}
\vskip 6pt
\centerline{\bf from Improved Lattice Actions}
\vskip2cm
\hskip4cm J.P.  Ma  and B.  H.  J.  McKellar\par
\vskip0.5cm
\hskip4cm School of Physics\par
\hskip4cm Research Center for High Energy Physics \par
\hskip4cm University of Melbourne \par
\hskip4cm Parkville, Victoria 3052\par
\hskip4cm Australia \par
\vskip2cm
{\bf\underbar{Abstract}}:\par
We use improved lattice actions for glue, light quarks and heavy quarks 
for which we use lattice NRQCD to compute hadron masses.  Our results are in good agreement with experiment, except for charmed hadrons.  It seems that charmed quarks are not well approximated as heavy quarks nor as light quarks.
\par\vskip15pt
\noindent 
PACS numbers: 11.15Ha, 12.38G, 14.40G and 14.40Gx 
\par\vfil\eject
\ \ \
\par\vfil\eject}
\baselineskip=15pt
\pageno=1
\noindent 
{\bf 1.  Introduction}
\vskip 15pt
A notorious feature of the perturbative theory of lattice QCD is that 
the perturbation series converges slowly at a moderate lattice spacing. 
Predictions based only on 
first two or three terms in the series are in most cases still not 
reliable. A recent study[1] shows that 
a partial reason for this is the large contribution
from tadpole diagrams, which are absent if one uses dimensional 
regularization instead of lattice regularization. Based on this
the method of tadpole improvement[1] for lattice perturbation series has been suggested. 
This has many practical implications
for studies of lattice QCD. Most importantly it makes the 
idea of improved actions practicable in simulations
of lattice QCD. The idea of improved actions was proposed 
a long time ago[2], to systematically improve lattice actions
using perturbation theory to remove the effect 
introduced by the finite lattice spacing $a$. If such an improvement 
really works, one can simulate lattice QCD on a coarse lattice 
with a lower cost in CPU time, but still obtain reliable 
results for physics in the continuum limit. Several simulations 
[3--6], where the tadpole
improvement is implemented,
already show that this is possible. 
\par
With an improved action for glue at the one-loop level and the tree-level 
improved action for NRQCD the mass spectrum of charmonium  
is obtained in [3] from 
lattices with a lattice spacing $a$ between 0.5fm and 0.2fm, 
and it agrees with experiment. Similar results were also obtained 
in other simulations[4], where 
the mass spectra of bottonium and the $B_c$ meson, in addition to that of
charmonium, were obtained. With an improved action 
for light quarks, 
the D234 action, the mass spectrum of light hadrons is 
calculated from simulations on coarse lattices[5], and good agreement
with experiment is found. In general improved actions 
are not unique, in [6] another improved action for light fermion, 
which
was proposed long time ago[7], was employed,  the mass spectrum of light hadrons
was also successfully calculated[6], where the gluonic action 
improved at tree-level was used. Unlike the widely used SW-action 
for light quarks[8], in which the effect of $O(a)$ is removed at tree-level, 
the actions used in these simulations are improved up to $O(a^2)$ at 
tree-level. 
\par 
In this work we will present our results for the mass spectrum of
light hadrons, strange hadrons, and  
hadrons containing a b- or c-quark. 
In calculating the mass spectrum we use the improved action for glue  
at the one-loop level and the action of [7] for light quarks. 
For the heavy   b- and c-quarks we employ the action of 
lattice NRQCD.  The actions for quarks are all improved 
at tree-level upto $O(a^2)$. We use the quenched approximation 
for light quarks. Two lattices are used with the size
$6^3\times 12$ and $8^3\times 16$. Their lattice spacings 
determined by the charmonium spectrum are 0.41fm and 0.23fm respectively. 
The whole calculation is performed 
on two UNIX workstations, which have computational power equivalent 
to that of a high-performance personal computer (like a Pentium Pro 200). 
Our paper is organized as follows:
In Sect.2 we introduce the actions used in this work.
Our results for light hadrons and strange hadrons are given 
in Sect.3. The results of c-flavored 
and b-flavored hadrons are given in Sect.4.  Sect.5 is the summary.  
\par
\vskip 20pt
\par\noindent
{\bf 2.  The Improved Actions}
\par\vskip15pt
\par\noindent
{\bf 2.1 The gluonic action:}
\par
We take the one-loop improved action for glue[9], where 
the action consists of plaquette, rectangle
and paralellogram terms and is accurate up to errors of $O(\alpha_s^2 a^2, a^4)$.  
Implementing tadpole improvement the action becomes[3]

  $$\eqalign{ S(U) &= \beta \sum_{pl} {1\over 3}{\rm Re 
   Tr} (1-U_{pl}) + \beta_{rt}\sum_{rt} {1\over 3}{\rm ReTr}
   (1-U_{tr}) +\beta_{pg} \sum_{pg}{1\over 3}{\rm ReTr}
    (1-U_{pg}), \cr 
  \beta_{rt}&= -{\beta\over 20 u_0^2}(1+0.4805 \alpha_s), \ \ \ 
   \beta_{pg}=-{\beta\over u_0^2}0.03325 \alpha_s, \cr
   u_0 &=({1\over 3}{\rm ReTr}\langle U_{pl}\rangle )^{1\over 4}, \ \ \ 
 \alpha_s =-{\ln({1\over 3}{\rm ReTr}\langle U_{pl}\rangle )\over 3.06839 }.  \cr}
   \eqno(2.1) $$

We used this action to generate gluonic configurations at 
$\beta=7.4$, on a lattice whose size is $8^3\times 16$,    
and also at $\beta=6.8$ with the size $6^3\times 12$. 
The parameter $u_0$ is determined by self-consistency. It is 
0.8631 and 0.8267 at $\beta=7.4$ and $\beta=6.8$
respectively. 
The pseudo heat bath method[10] was used 
to update the links, and the three 
$SU(2)$ subgroups were updated 3 times in each 
overall update step.  
For each lattice we generated 100 configurations 
for our mass calculations.  
\par
\vskip5pt
\noindent
{\bf 2.2 The improved action for light quarks:}
\par
The improved action proposed in [7] can be written as:
  $$ S_{\rm light} = 
    -\sum_x \left\{
    m\bar\psi (x) \psi(x) + \sum_\mu \bar\psi (x) 
    \gamma_\mu \Delta_\mu (1-c_1\Delta^{(2)}_\mu) \psi (x) 
    +r\sum_\mu \bar\psi (x) \Delta^{(2)}_\mu\Delta^{(2)}_\mu 
    \psi (x) \right \} , \eqno(2.2) $$ 
where $\Delta_\mu$ and $\Delta^{(2)}_\mu$ are lattice 
derivatives with the gauge link $U_\mu(x)$: 
  $$ \eqalign {  \Delta_\mu \psi(x)&={1\over 2} ( 
      U_\mu (x) \psi(x+\hat\mu) -U^\dagger_\mu (x-\hat\mu)
      \psi (x-\hat\mu)), \cr  
      \Delta_\mu^{(2)} \psi(x) &= 
       U_\mu (x) \psi(x+\hat\mu)+U^\dagger_\mu (x-\hat\mu)
      \psi (x-\hat\mu)-2\psi(x).\cr}  \eqno(2.3)$$
The parameter $c_1$ is determined to be 1/6 at tree-level to 
remove the effect at $O(a^2)$, $m$ is the mass parameter 
for a quark. The last term in the action with the parameter $r$  
is introduced in analogy to the Wilson term in the Wilson action 
to solve the doubling problem of lattice fermions. However,
the doubling problem is not totally solved. In the free case one can solve 
the equation of motion determined by the action to see
whether the doublers are removed or not. An analysis in [8], 
and the analysis in [5] for the D234 action, show that in the low-energy
regime the actions describe one particle in the sense that 
the propagator has only one pole. In high-energy regime
there are additional ``unphysical'' poles. As we are only interested 
in the low-energy regime as we are in this work, we can expect 
that the effect from other ``unphysical poles'' to be negligible. 
We will take $r=1/6$. With this choice the action is the same as employed 
in [6], in which it is shown that there is indeed no effect
in the low-energy regime 
which can be related to the ``unphysical poles''. 
In this work we used the stabilized biconjugate 
gradient algorithm[11] to calculate propagators 
for light quarks. On our lattices this algorithm is at least 
three times faster than   
the conventional conjugate gradient algorithm. 
\par\vskip10pt
\noindent
{\bf 2.3 The Action of Lattice NRQCD}
\par
Heavy quarks whose mass is larger than 1 in lattice units 
cannot be simulated directly as above with reliable
results. To simulate them one uses the heavy quark effective theory, HQET. 
The formulation of HQET for hadrons with 
zero velocity is equivalent to that of NRQCD on the lattice, except that 
the expansion parameters are different. As we will only create 
hadronic states on lattice with zero
space-momenta, 
we may use lattice NRQCD for heavy quarks like the b- and c-quarks. 
On the lattice one needs to calculate 
the propagator of heavy quarks satisfying 
its appropriate evolution equation.  We take the evolution equation 
proposed in [12,13].  The propagator $G(t)$ 
(where $G(t)=0$ for $t\le 0$) can be calculated on the lattice
as:  
$$ \eqalign {G(1) &=\biggl(1-{H_0\over 2n}\biggr)^n U_4^\dagger
     \biggl(1-{H_0\over 2n}\biggr)^n \delta_{\bf {x},0}, \cr
    G(t+1) & = \biggl(1-{H_0\over 2n}\biggr)^n U_4^\dagger
     \biggl(1-{H_0\over 2n}\biggr)^n (1-\delta H) G(t), \cr} \eqno(2.4)$$
where 
  $$ \eqalign { 
    H_0 &= -{\Delta ^{(2)}\over 2 M_Q}, \cr  
    \delta H &=-{ g\over 2M_Q }
         {\bf\sigma\cdot B}+{  \Delta^{(4)}\over 24M_Q}- 
    { (\Delta ^{(2)})^2\over
    16nM_Q^2} .  \cr }   \eqno(2.5)$$ 
In (2.5) $\Delta^{(4)}$
is the lattice version of the continuum operator $\sum_i D_i^4$.
$M_Q$ is the mass parameter for the heavy quark Q.   
The last two terms in $\delta H$ are the correction
terms to remove the effect at order $O(a^2)$. 
The first term is responsible for spin-splitting in the mass spectrum, 
where ${\bf B}$ is the chorommagnetic field. We use definition of  
 ${\bf B}$ in terms of gauge links given in [12], and it is also 
improved up to errors of order  $O(a^4,g^2a^2)$. 
The parameter $n$ is introduced to avoid numerical 
instability when high-momentum modes occur.  
With propagators calculated with Eq.(2.4) we reach an accuracy
 of order ${1\over M_Q}$ in the mass 
spectrum. 
With lattice NRQCD we have calculated the mass spectrum of quarkonium 
at the above $\beta$-values[4], and have determined that at $\beta =7.4$ 
the mass parameters for $b$- and $c$-quarks are $M_b=4.6$ and $M_c=1.4$ 
respectively.  
We will use 
these parameters for our calculations of mas spectra of b- and c-flavored
hadrons. We take $n=3$ for c-quark and $n=1$ for b-quark. 
\par
In our calculations of quark propagators with the action in Eq.(2.2) or with 
Eq.(2.4) tadpole improvement is implemented.                
\par\vskip20pt
\noindent
{\bf 3 The Mass of Light Hadrons }
\par\vskip20pt
We create hadronic states on the lattice by using standard local operators
$O_H(x)$:
 $$ \eqalign {
      O_\pi (x)&= \bar u (x) \gamma_5 d(x), \cr
      O_\rho^\mu (x)&=\bar u(x) \gamma^\mu d(x), \cr
      O_P(x) &= \varepsilon_{abc} u_a(x) \left [ 
       u_b^T (x) C \gamma_5 d_c(x)\right ] \cr} \eqno(3.1)$$ 
for $\pi$, $\rho$ and proton respectively. In Eq.(3.1) $u(x)$ and $d(x)$
stand for $u$- and $d$-quarks, the indices $a,b$ and $c$ are color indices. 
With these operators one can measure the corresponding 
hadron correlations $C_H(t)$, in which the hadron has zero space-momentum 
and its spin is averaged if it has spin. We only use local sources to 
calculate $C_H(t)$, and $u$- and $d$-quarks are taken to be degenerate in mass. 
The hadron correlation $C_H(t)$ is fitted in a certain time-interval as:
   $$\eqalign { a_H& \left ( e^{-m_H t}+e^{-m_h(T-t)}\right ) 
   \ \ \ \ {\rm for} \ H=\pi, \ \rho, \cr 
          a_H& e^{-m_H t}  \  {\rm with } \ t< {T\over 2}, 
   \ \ \ {\rm for \ H=proton}, \cr} \eqno(3.2)$$  
where $T$ is the lattice size in the time-direction. 
\par
Using the lattice configurations on an $8^3\times 16$ lattice for $\beta =7.4$ 
we calculated these hadron correlations varying the mass parameter for light
quark from $-0.5$ to $-0.75$, and fit them according to Eq.(3.2).
The negative values of the mass parameters simply reflect the 
difference in sign between the mass parameter and the ``Wilson'' parameter $r$.  The range of 
light quark mass parameters investigated here are the same order as
the mass of the $s$-quark. 
The fitting window for $C_\pi(t)$ and 
$C_\rho (t)$ is chosen from $t=6$ to $t=10$, for the proton we can see 
a plateau in the region of $t=4$ to $t=8$, so we take the window 
to be from $t=4$ to $t=7$. As examples which show our fit, we plot 
the measured $C_H(t)$ 
for $\pi$ and $\rho$  with the fitted results in Fig.1A 
and Fig.1B respectively. Fig.1C is the plot for the effective mass
of the proton. In all these figures the results are at $m=-0.65$.   
Our results of fitted masses are given in Table 1.  
\par\vskip10pt
\centerline{\bf Table 1} 
\par\vskip10pt
\vbox{\tabskip=0pt \offinterlineskip
\def\tablerule{\noalign{\hrule}}
\halign to400pt{\strut#& \vrule#\tabskip=1em plus2em&
 \hfil#& \vrule #& \hfil#&\vrule#&\hfil#&\vrule#&\hfil#& \vrule#&\hfil#&
 \vrule#&\hfil#&\vrule#
\tabskip=0pt\cr\tablerule
 && $\beta=7.4$  && $m=-0.5$  && $m=-0.6$  &&$m=-0.65$&&$m=-0.7$ && $m=-0.75$
                                                             &\cr\tablerule
&& $m_\pi$  && 1.142(9) &&  0.941(11) &&  0.832(14)
   && 0.713(18)  && 0.577(27)   &\cr\tablerule
&& $m_\rho$  && 1.362(20)  &&  1.211(24) && 1.138(32)
  && 1.068(46)  &&  1.010(95)        &\cr\tablerule
&& $m_P$  && 2.23(9)  &&  1.96(7)  && 1.82(6)
   && 1.69(7)  &&  1.56(16)        &\cr\tablerule
 \hfil\cr}}
\par
We also 
constructed hadron correlations for pion with 
a minimal lattice momentum $\vert {\bf p} \vert ={ 2\pi\over L}$, where 
$L$ is the lattice size in the space-direction. Measuring these 
correlations one can investigate the dispersion relation 
$E^2({\bf p})-m^2=c^2\vert {\bf p}\vert^2$.  
We find that $c$ is close to 1. For example, we obtained 
$c=1.02(7)$ at $m=-0.5$ and $c=1.00(18)$ at $m=-0.75$. 
This fact indicates  
that 
rotational invariance is restored on the lattice.  
To find the masses in the chiral limit, we fit the hadron masses with:
   $$ m_\rho =v_0 +v_1 m^2_\pi, \ \ \, m_P=u_0 +u_1m_\pi^2 +
     u_2 m_\pi^3. \eqno(3.3) $$ 
We find that the masses given in Table 1 are well described 
by these relations as  shown in Fig.2. For the mass of the proton 
we clearly see the effect from the cubic term. With these fits 
one can extrapolate the masses to the chiral limit. 
\par
If we assume that the quark mass is obtained through 
multiplicative and additive renormalizations of the mass parameter, 
then the square of $m_\pi$ should be linear in the mass parameter $m$. 
With the data in Table 1 we find  good agreement with this relation. 
Hence we are able to determine the critical mass-parameter $m_0$ 
at which $m_\pi=0$.  We also determine the mass parameter $m_s$ of
the strange quark from the experimental value $m_K^2/m_\rho^2=0.412$. 
These parameters at $\beta=7.4$ are: 
  $$ m_0=-0.830(40), \ \  m_s=-0.6658. \eqno(3.4)$$
With the $m_s$ above we calculated the propagator of the s-quark 
and then the hadron correlations for $K$, $K^*$ and $\phi$, 
where $C_K(t)$ and $C_{K^*}(t)$ are calculated with these mass parameters
in Table 1 for light quark. The masses obtained for $m_K$, $m_{K^*}$  
with non-zero mass for the $u$-and $d$-quark are given in Table 2. 
\par\vfil\eject
\par\vskip 10pt
\centerline{\bf Table 2}
\par\vskip10pt
\vbox{\tabskip=0pt \offinterlineskip
\def\tablerule{\noalign{\hrule}}
\halign to400pt{\strut#& \vrule#\tabskip=1em plus2em&
 \hfil#& \vrule #& \hfil#&\vrule#&\hfil#&\vrule#&\hfil#& \vrule#&\hfil#&
 \vrule#&\hfil#&\vrule#
\tabskip=0pt\cr\tablerule
 && $\beta=7.4$  && $m=-0.5$  && $m=-0.6$  &&$m=-0.65$&&$m=-0.7$ && $m=-0.75$
                                                             &\cr\tablerule
 &&$m_K$ &&  $0.979(12) $  && 0.870(13) &&  0.814(14) && 0.755(16)
   && 0.694(20)  &\cr\tablerule
&& $m_{K^*}$  && 1.239(25)  &&  1.163(29) && 1.127(33)
   && 1.092(40)  && 1.061(50)  &\cr\tablerule
   \hfil\cr}}
\par 
We fit these masses with: 
  $$ m_K^2 =b_0 +b_1(m-m_0), \ \ \ m_{K^*} =c_0 +c_1(m-m_0) \eqno(3.5)$$
to extrapolate to the limit of zero-mass  $u$- and $d$-quarks. 
The data and the fitted line are drawn in Fig.3. The data points are 
well described by the linear fitted line giving confidence in the extrapolation.
\par 
Doing the same calculations of $C_H(t)$ with our configurations
at $\beta=6.8$, for which we choose the region of the light quark mass 
parameter $m$ from $-0.8$ to $-1.1$, we obtain hadron masses which are 
given in Table 3. 
\par\vskip 10pt
\centerline{\bf Table 3.} 
\par\vskip10pt
\vbox{\tabskip=0pt \offinterlineskip
\def\tablerule{\noalign{\hrule}}
\halign to400pt{\strut#& \vrule#\tabskip=1em plus2em&
 \hfil#& \vrule #& \hfil#&\vrule#&\hfil#&\vrule#&\hfil#& \vrule#&\hfil#&
 \vrule#&\hfil#&\vrule#
\tabskip=0pt\cr\tablerule
 && $\beta=6.8$  && $m=-0.8$  && $m=-0.9$  &&$m=-1.0$&&$m=-1.05$ && $m=-1.1$
                                                             &\cr\tablerule
 && $m_\pi$  && 1.353(9) &&  1.179(10) &&  0.986(11)
   && 0.877(13) &&  0.754(14)    &\cr\tablerule
&& $m_\rho$  && 1.703(21)  &&  1.587(29) && 1.470(48)
   && 1.409(62)   &&  1.340(81)        &\cr\tablerule
&& $m_N$  && 2.770(77)  &&  2.523(77)  && 2.262(79)
   && 2.113(86)   &&  1.941(106)         &\cr\tablerule
  &&$m_K$ &&  $1.218(10) $  && 1.125(10) &&  1.028(11) && 0.977(12)
   && 0.924(12)  &\cr\tablerule
&& $m_{K^*}$  && 1.614(27)  &&  1.554(33) && 1.495(43)
   && 1.467(49)  && 1.439(56)  &\cr\tablerule
\hfil\cr}}
\par\vskip10pt
At this $\beta$-value we have determined: 
 $$ m_0=-1.233(56), \ \ \ \ m_s=-0.9513. \eqno(3.6)$$ 
With our results we obtain hadron masses in the chiral limit and
the mass $m_\phi$ which depends only on $m_s$.  They are given in Table 4
as a ratio to  $m_\rho$,which is itself given  in lattice units.
\par\vskip 10pt
\centerline{\bf Table 4}
\vskip 10pt
\vbox{\tabskip=0pt \offinterlineskip
\def\tablerule{\noalign{\hrule}}
\halign to400pt{\strut#& \vrule#\tabskip=1em plus2em&
 \hfil#& \vrule #& \hfil#&\vrule#&\hfil#&\vrule#&\hfil#&
 \vrule#
\tabskip=0pt\cr\tablerule
 && \ \   && $\beta =6.8$   && $\beta =7.4$  && Exp. &\cr\tablerule
 && $am_\rho$  &&  1.198(67) && 0.885(45) && \ \  & \cr\tablerule
&& $m_P/m_\rho $  && 1.16(10) &&  1.45(11)  && 1.22 &\cr\tablerule
&& $m_K/m_\rho $  && 0.63(4) && 0.64(3)  && 0.64 &\cr\tablerule
&& $m_{K^*}/m_\rho $  && 1.14(7)  &&  1.13(6)  && 1.16 &\cr\tablerule
&& $m_\phi /m_\rho $  && 1.27(8)  &&  1.26(9)  && 1.32  &\cr\tablerule
\hfil\cr}}
\par
Having chosen the parameter $m_s$ to fit the  
mass of the $K$, our predictions for the meson masses
are in good agreement with experiment, although the quenched approximation 
was used. The mass of proton obtained at $\beta=7.4$ is  
$20\%$ larger than the experimental value, while that from 
$\beta =6.8$ agrees with the experimental value 
within the statistical error.  
A possible reason for the deviation at $\beta =7.4$, apart from  
the quenched approximation, could be the effect of the finite volume. 
Another possible reason is the use of a local source in our work, 
a result of which is that 
the signal of the ground state in $C_P(t)$ 
is rather weak, and the fitting quality of $C_P(t)$
is worse than that of meson propagators. This situation may be improved
if smeared sources are used.  
\par
From the calculated $\rho$-mass in lattice units we can determine  
the physical lattice
spacing. We obtain:
  $$ \eqalign {  a_\rho^{-1}&=0.64(4){\rm GeV,\ \ \ at\ } \beta =6.8, \cr 
         a_\rho^{-1}& =0.87(4){\rm GeV,\ \ \ at\ } \beta =7.4.
        \cr} \eqno(3.7) $$
It is instructive to compare the lattice spacings determined 
from the quarkonium system. In our previous work[4] we obtained:
  $$ \eqalign { a_{J/\Psi}^{-1}&=0.477(5){\rm GeV}  
     \ \ \ \ \ \ \ \ \ \ \ \ {\rm at\ } \beta=6.8, \cr     
a_{J/\Psi}^{-1}& =0.749(4){\rm GeV}, \ \ \  a_{\Upsilon}^{-1} 
    =0.861(5){\rm GeV}\ \ \  {\rm at\ } \beta=7.4. \cr}  \eqno(3.8)$$  
The lattice spacing at $\beta=7.4$ determined from $\rho$ is close to that 
obtained from 
$\Upsilon$, while there is a deviation of $14\%$ in comparison 
with $a_{J/\Psi}^{-1}$. At $\beta =6.8$ the deviation is at $30\%$. 
Similar lattice spacings were also obtained with D234 action in [5].
Possible reasons for these deviations include:  
the effects neglected in the quenched approximation being  
different in different systems,  
the effect from higher orders in lattice NRQCD  
can 
be large for charmonium system because the charm mass is not 
enough large. 
In general one should not be surprised that the lattice spacings determined from   
different systems are  different when approximations 
are used in the calculations.       
\par 
From the determined dependence of $m_\rho$ on $m_\pi$ we can obtain 
the quantity $J=m_\rho dm_\rho^2 /dm_\pi^2$ at the experimental value 
$m_\rho /m_\pi=1.8$. 
This quantity is introduced in [14] to judge the quality of the quenched 
approximation. For the ``real world'' it is $J=0.48(2)$, while 
the ``world averaged'' result from the quenched approximation is $J=0.37$. 
From our results we obtain:
  $$ J=0.375(87) \ {\rm at\ } \beta=6.8,\ \ \ \ 
     J=0.363(63) \ {\rm at\ } \beta=7.4, \eqno(3.9) $$
which are in consistent with $J=0.37$. 
\par\vskip20pt
\bigbreak
\noindent
{\bf 4. The Mass Spectrum of c- and b- flavored Hadrons}
\par\vskip20pt
In this section we will present only our results at $\beta =7.4$
in detail. At $\beta =6.8$ there is  a problem for simulations of 
$b$-quark with lattice NRQCD. We find large correlation lengths,
 so that a significant
finite-volume effect should be expected. 
On this lattice we calculated only masses of c-flavored hadrons and 
we will mention these results briefly. With the 
configurations at $\beta =7.4$ we will calculate masses of the hadrons  
$B^0,\ B^*, \ B_s, \ B_s^*$  and $\Lambda_b$ and masses of the  
$D^0,\ D^*, \ D_s, \ D_s^*$  and $\Lambda_c$.  
\par
For the hadrons we use operators analogous to the operators 
listed in Eq.(3.1) 
to create these hadronic states on lattice.  
The difference is that a 
light quark field in Eq.(3.1) is replaced by a heavy quark field $Q(x)$
and the field $Q(x)$ only has two nonzero components. 
The propagation of the field is determined by Eq.(2.4). For
light quarks we use the propagators of the last section.
For calculating propagators of heavy quarks 
we use a source smeared by a gaussian:
  $$    F({\bf x}, {\bf y},t)  =\left [ 
     1+ \epsilon \sum_i \Delta ^{(2)}_i \right ]^{n_s}
     ({\bf x}, {\bf y},t). \eqno(4.1) $$ 
The smearing parameter $n_s$ is fixed at $n_s=10$, while 
the parameter $\epsilon$ is adjusted so that the smearing radius 
is about the half of the simulated system. Using a smeared source 
is essential for our simulations. With a local source for heavy quark 
propagators the signal of hadron correlations is overwhelmed 
by the statistical noise after 3 or 4 time slices from the source. 
We only calculate hadron correlations with smeared sources,  
while the sinks remain unsmeared. 
The measured hadron propagators are fitted to the form
  $$  C_H(t) \sim a_H e^{-E_H t},  \eqno(4.2)$$ 
where the index $H$ refers to the hadrons mentioned above. With 
the fitted energy $E_H$ one can obtain the hadron mass $M_H$
   $$M_H =\Delta_Q +E_H,     \eqno(4.3)$$
where $\Delta_Q$ is the difference between the renormalized mass
of the heavy quark and the zero point energy of the heavy quark on lattice,
and it does not depend on the type of hadron. Therefore one can predict 
the mass difference between two types of hadrons if the $E_H$'s are known. 
\par
Although we used a smeared source for heavy quark propagator, 
the signal is buried in statistical noise after 10 or 11 time slices
from the source for mesons and after 8 or 9 time slices for baryons. 
For the meson we already see a plateau in the effective mass. We will take 
the fit-window for meson propagators from 6 to 9. For baryons  
we take the fit-window 
from 5 to 7. In Fig.4a, Fig. 4b and Fig. 4c we give our effective mass 
plots for $B$, $B^*$ and $\Lambda_b$ respectively. These 
plots are for propagators at the mass parameter $m=-0.65$. 
Our results for $E_H$ of various hadrons with nonzero mass of 
light quarks are given in Table 5.

\par\vskip10pt
\centerline{\bf Table 5} 
\par\vskip10pt
\vbox{\tabskip=0pt \offinterlineskip
\def\tablerule{\noalign{\hrule}}
\halign to400pt{\strut#& \vrule#\tabskip=1em plus2em&
 \hfil#& \vrule #& \hfil#&\vrule#&\hfil#&\vrule#&\hfil#& \vrule#&\hfil#&
 \vrule#&\hfil#&\vrule#
\tabskip=0pt\cr\tablerule
 && \ \  && $m=-0.5$  && $m=-0.6$  &&$m=-0.65$&&$m=-0.7$ && $m=-0.75$
                                                             &\cr\tablerule
 && $E_{D^0} $  && 1.069(10) && 0.990(10)  &&
  0.950(9) && 0.909(9) && 0.863(9)     &\cr\tablerule
&& $E_{D^*} $  && 1.143(8)  && 1.067(9)  && 1.030(9)
   && 0.991(10)  && 0.948(10)    &\cr\tablerule
&& $E_{\Lambda_c} $  && 1.976(16)  && 1.810(17) &&  1.727(17)
   && 1.643(15) &&  1.561(17)    &\cr\tablerule
 && $E_{B^0} $  && 1.043(14) &&  0.967(12) && 0.928(12)
   && 0.894(12) &&  0.856(13)    &\cr\tablerule
&& $E_{B^*} $  && 1.066(11)  && 0.999(9)  && 0.962(9)
   && 0.927(9)  &&  0.891(10)   &\cr\tablerule
&& $E_{\Lambda_b} $  && 1.892(12)  && 1.713(12)  &&  1.617(14)
   &&  1.504(16) && 1.381(22)     &\cr\tablerule
  \hfil\cr}}
\par\vskip10pt
\par\vskip10pt
For c- and b-flavored hadrons with s-quarks we have:
  $$ \eqalign { E_{B_s}&=0.920(11), \ \ \ E_{B_s^*}=0.952(9), \cr 
                E_{D_s}&=0.937(9), \ \ \  E_{D_s^*}=1.018(9). \cr}
  \eqno(4.4)$$ 
To extrapolate to the zero light quark mass we assume the dependence 
of $E_H$ on the mass parameter $m$ to be:
  $$ E_H = E_H^{(0)} +E_H^{(1)} (m-m_0), \eqno(4.5) $$
and use this relation to fit our data. 
In Fig.5A and Fig.5B we show the fits to the data of Table 5.
Except for $\Lambda_b$, our results in Table 5 can be well 
fitted with this relation. 
In Fig.5A the fitted line
for $\Lambda_b$  
is determined only by the first 4 data points. The fit has  
$\chi^2=0.76$. If we include the last point into the fit, 
the $\chi^2$ jumps to 3.39, and $E_{\Lambda_b}^{(0)}$ changes 
from 1.218(7) to 1.250(9). A reason for
this may be the effect of higher orders of $(m-m_0)$ omitted in Eq.(4.5). 
In our final results 
we will use the fit with the 4 data point for $\Lambda_b$. 
We obtain the energies at zero mass of light quarks: 
$$ \eqalign { 
     E_{B^0}^{(0)} &=0.796(4), \ \ \ E_{B^*}^{(0)} =0.836(3), \ \ \  
     E_{\Lambda_b}^{(0)}=1.218(7), \cr
     E_{D^0}^{(0)} &=0.800(3), \ \ \ E_{D^*}^{(0)} =0.889(3), \ \ \  
     E_{\Lambda_c}^{(0)}=1.428(5). \cr} \eqno(4.6) $$
\par
It is interesting to note that the energies given in Eq.(4.4) and 
in Eq.(4.6) for b-flavored hadrons are close to 
those 
of c-flavored hadrons. 
These energies $E_H$'s can be expanded in the inverse of $M_Q$ with a leading 
order $M_Q^0$ and are calculated with an accuracy of $M_Q^{-1}$ 
in this work. 
The above fact indicates that the effect from the next-to-leading order  
and from higher orders is small in these spin averaged masses. However, as we will 
see, this is not true for spin-splittings.    
With the results in Eq.(4.4) and in Eq.(4.6) 
we are able to predict the mass differences. 
For doing this we take the lattice spacing determined by the  
bottonium system for b-flavored hadrons, 
$$ a_b^{-1} =0.86 {\rm GeV}. \eqno(4.7) $$
This spacing is the same
within errors    
as the lattice spacing determined from $m_\rho$ in the last section. 
We obtain
  $$ \eqalign {
    M_{B^*}-M_{B^0}& =34(6){\rm MeV}, \ \ \
   M_{B_s}-M_{B^0}=107(13){\rm MeV},\cr   
     M_{B_s^*} -M_{B_s}&=27(17){\rm MeV},  \ \ \
   M_{\Lambda_b}-M_{B^0}=363(9){\rm MeV}.\cr} \eqno(4.8) $$
These results should be compared with the experimental results
  $$ \eqalign {
   \ \ \ M_{B^*}-M_{B^0}&=46{\rm MeV}, \ \ \
   M_{B_s}-M_{B^0}=91{\rm MeV},\cr   M_{B_s^*} -M_{B_s}&=47{\rm MeV},  \ \ \
   M_{\Lambda_b}-M_{B^0}=363{\rm MeV}.\cr}   \eqno(4.9) $$    
We find that our results for $M_{\Lambda_b}-M_{B^0}$ and $M_{B_s}-M_{B^0}$
agree well with experimental results, while the spin-splitting 
$M_{B^*}-M_{B^0}$ and $M_{B_s^*} -M_{B_s}$ are not in such good agreement, 
but differ by less than 2 standard deviations from the observed values. 
The value 
of $M_{B^*}-M_{B^0}$ is $28\% $ 
lower than experimental value, but it agrees with 
the result from a large and fine lattice[15].
\par
For c-flavored hadrons we take the lattice spacing determined by the  
charmonium system: 
  $$ a_c^{-1}=0.75{\rm GeV}  \eqno(4.10) $$
and obtain
  $$ \eqalign {
    M_{D^*}-M_{D^0}&=67(5){\rm MeV}, \ \ \
   M_{D_s}-M_{D^0}=103(9){\rm MeV},\cr   M_{D_s^*} -M_{D_s}& =66(14){\rm MeV},  \ \ \
   M_{\Lambda_c}-M_{D^0}=471(6){\rm MeV}.\cr}   \eqno(4.11) $$
The experimental results are: 
  $$ \eqalign {
  M_{D^*}-M_{D^0}&=143{\rm MeV}, \ \ \
   M_{D_s}-M_{D^0}=104{\rm MeV},\cr   
   M_{D_s^*} -M_{D_s}&=144{\rm MeV},  \ \ \
   M_{\Lambda_c}-M_{D^0}=420{\rm MeV}.\cr}   \eqno(4.12) $$
Our value for $M_{D_s}-M_{D^0}$ agrees well with the experimental value, 
but the predicted $M_{\Lambda_c}-M_{D^0}$ is $12\%$ larger the experimental 
value. More worse are the spin-splittings which are  much lower 
than experimental values. Such discrepancies may be expected as
we used lattice NRQCD upto order of $1/M_Q$. For $c$-quarks the effect 
from higher orders is very significant because the charm mass 
is not very large. A recent study of quarkonium systems[16]  
also shows that the effect of higher order terms in $1/M_Q$ is large.    
We also have calculated masses of c-flavored hadrons 
at $\beta =6.8$ with  similar results.   
It shows that the action for  lattice NRQCD employed here is not accurate enough 
for calculations of  the spin-splitting for c-flavored hadrons. 
\par
To obtain the absolute mass of a b- or c-flavored hadron we need to know 
$\Delta_Q$ in Eq.(4.3). This quantity can be calculated perturbatively. 
It can also be extracted by studying the mass spectrum of the quarkonium system. 
We extracted this number from our previous study at $\beta=7.4$: 
  $$  \Delta_c=1.59(11), \ \ \ \Delta_b=5.25(18). \eqno(4.13)$$
It should be noted that effects from the spin-dependent interaction
are neglected in the extraction. 
Adding 
the spin-dependent interaction has little effect on these numbers, 
and also these numbers have a quite large uncertainty. Using  
Eq.(4.13) we obtain  
  $$ M_{D^0}=1.79(8){\rm GeV},\ \ \ \ \ 
        M_{B^0}=5.20(15) {\rm GeV} \eqno(4.14)$$
where the lattice spacings in Eq.(4.10) and in Eq.(4.7) are used respectively. 
These predictions are in good agreement with the experimental values: 
 $$ M_{D^0}=1.864{\rm GeV }, \ \ \  M_{B^0}=5.278 {\rm GeV}. \eqno(4.15) $$
However, the accuracy of our predictions is not good because of 
  the large statistical error of   
$\Delta_Q$.
\par\vskip20pt
\noindent
{\bf 5. Summary}
\par\vskip 20pt
In this work we used improved actions for glue and quarks to calculate 
masses of  light, strange, c- and b- flavored 
hadrons, where for c- and b-quarks we employed lattice NRQCD.  
The actions for quarks are improved at tree-level to remove 
the effect at order of $O(a^2)$, while the gluonic action is 
improved at one-loop level. Tadpole improvement is implemented. 
The results obtained in this work for light and strange mesons  
are in agreement with experimental 
results. We obtain the mass spectrum of b-flavored hadrons 
with improved actions on a coarse lattice with the lattice spacing as 
0.23fm, and it is in agreement with  experiment.  
For c-flavored hadrons the large effect from higher orders neglected in 
lattice NRQCD prevents us from obtaining  spin-splitting in mass comparable with 
experiment. 
In this work we have shown that improved actions with tadpole
improvement can be used not only in light hadron sectors and in quarkonium 
systems as shown already in previous studies, but also works in the 
case of hadrons containing one heavy quark. 
\par\vskip 30pt
\noindent
{\bf Acknowledgment:}
\par
This work is supported by Australia Research Council. We thank our 
computer manager Dr. M. Munro for help with the implementation of our
calculations. 
\par
\vfil\eject
\vskip 30pt
\centerline{\bf Reference} \par
\vskip20pt
\noindent 
[1] G.P.  Lepage and P.B.  Mackenzie, Phys.  Rev.  D48 (1993) 2250
\par\noindent
[2] K.  Symanzik in Mathematical Problems in Theoretical Phyics, 
ed.  R.  Schrader et al.  
\par\noindent \ \ \ \ Lecture Notes in Physics 153 (Springer 
Berlin 1982) 
\par\noindent
\ \ \ \ Nucl.  Phys.  B226 (1983) 187   
\par\noindent
[3] M.  Alford et.  al.  Phys.  Lett.  B361 (1995) 87
\par\noindent
[4] J.P. Ma and B.H.J. McKellar, Melbourne Preprint UM-P-96/55 
\par\noindent
[5] M. Alford, T. Klassen and G.P. Lepage, Nucl. Phys.(Proc. Suppl.)
B47 (1996) 370 
\par\noindent
[6] H.R. Fiebig and R.M. Woloshyn, Phys. Lett. B385 (1996) 273 
\par\noindent
[7] H. Hamer and C.M. Wu, Phys. Lett. B133 (1983) 351
\par\noindent
\ \ \ \ W. Wetzel, Phys. Lett. B136 (1984) 407
\par\noindent
\ \ \ \ T. Eguchi and N. Kawamoto, Nucl. Phys. B237 (1984) 609
\par\noindent
[8] B. Sheikholeslami and R. Wohlert, Nucl. Phys. B259 (1985) 572
\par\noindent
[9] M.  L\"uscher and P.  Weisz, Comm.  Math.  Phys.  97 (1985) 59
\par\noindent 
[10] N.  Cabibbo and E.  Marinari, Phys.  Lett.  B119 (1982) 387
\par\noindent
[11] A. Frommer et.al. Int. J. Mod. Phys. C5 (1994) 1073
\par\noindent
[12] G.P.  Lepage et al, Phys.  Rev.  D46 (1992) 4052
\par\noindent
[13] C.T.H.  Davies et al, Phys.  Rev.  D50 (1994) 6963
\par\noindent
[14] P. Lacock and C. Michael, Phys. Rev. D52 (1995) 5213 
\par\noindent 
[15] A. Ali Khan et al., Phys. Rev. D53 (1996) 6433 
\par\noindent
[16] H.D. Trottier, Preprint SFU HEP-131-96, hep-lat 9611026  
\par\noindent
\par\vfil 
\eject
\centerline{\bf Figure Caption}
\par\vskip20pt
\noindent
Fig.1A: The propagator $C_\pi(t)$ with $m=-.65$. 
The points with error are measured, 
the line is from the fit. The x-axis is for $t$. 
\par
\noindent
Fig.1B: The propagator $c_\rho (t)$ with $m=-.65$.
\par\noindent
Fig.1C: The effective mass plot for the proton propagator with $m=-.65$,   
the effective mass is defined as $m_{\rm eff}=\ln ( C_p(t-1)/C(t))$. 
The x-axis is for $t$. 
\par\noindent
Fig.2: The chiral extrapolation for $\rho$ and proton. The upper line
with data points is for proton. The lower line with data points is for
$\rho$. The x-axis is read to be $m_\pi^2$. 
\par\noindent
Fig.3: The chiral extrapolation for strange meson. 
\par\noindent
Fig.4A: The effective mass plot for $B$ with $m=-0.65$. 
\par\noindent
Fig.4B: The effective mass plot for $B^*$ with $m=-0.65$.
\par\noindent
Fig.4C: The effective mass plot for $\Lambda_b$ with $m=-0.65$.
\par\noindent
Fig.5A: The chiral extrapolation for b-flavored hadrons. 
The upper line with data points is for $\Lambda_b$, 
the middle one with data points is for $B^*$, 
the lower one with data points is for $B$. The x-axis is 
read as $m$. 
\par\noindent
Fig.5B: The chiral extrapolation for c-flavored hadrons. 
The upper line with data points is for $\Lambda_c$, 
the middle one with data points is for $D^*$, 
the lower one with data points is for $D$. The x-axis is 
read as $m$. 
\par\vfil\eject 
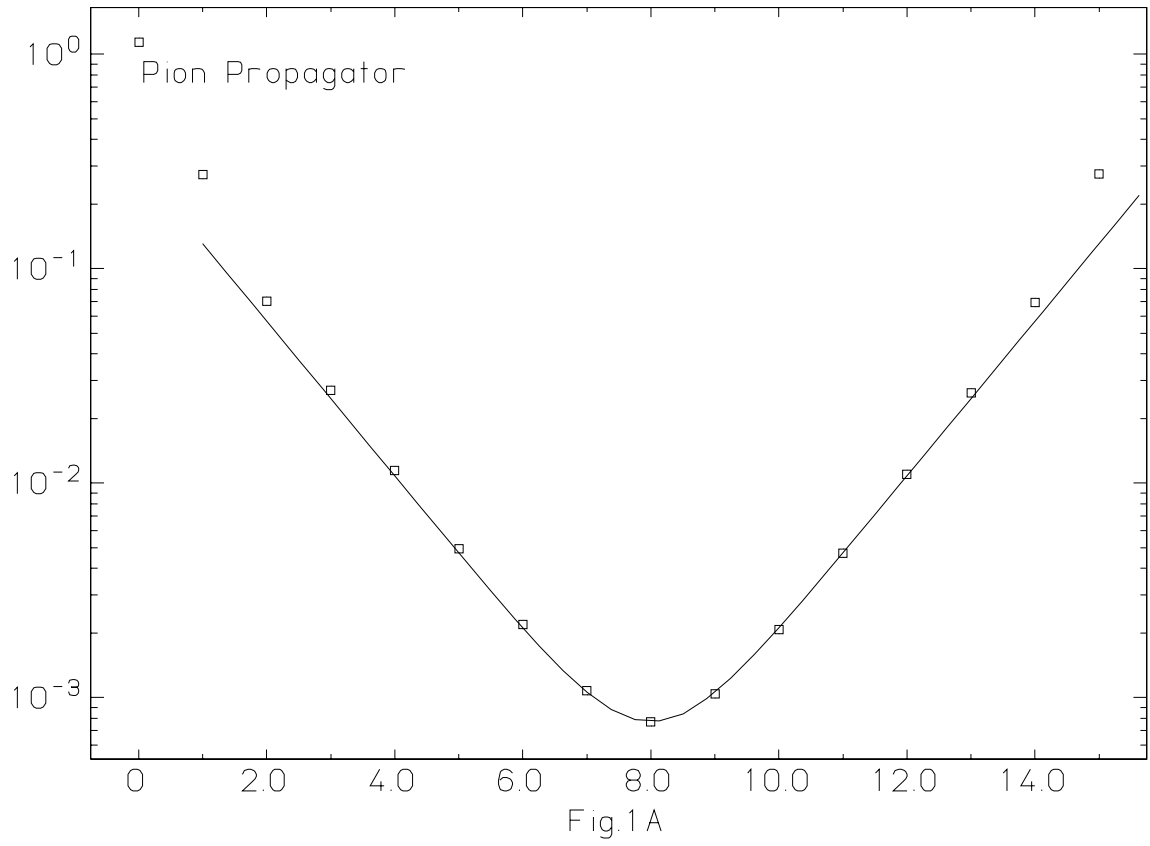
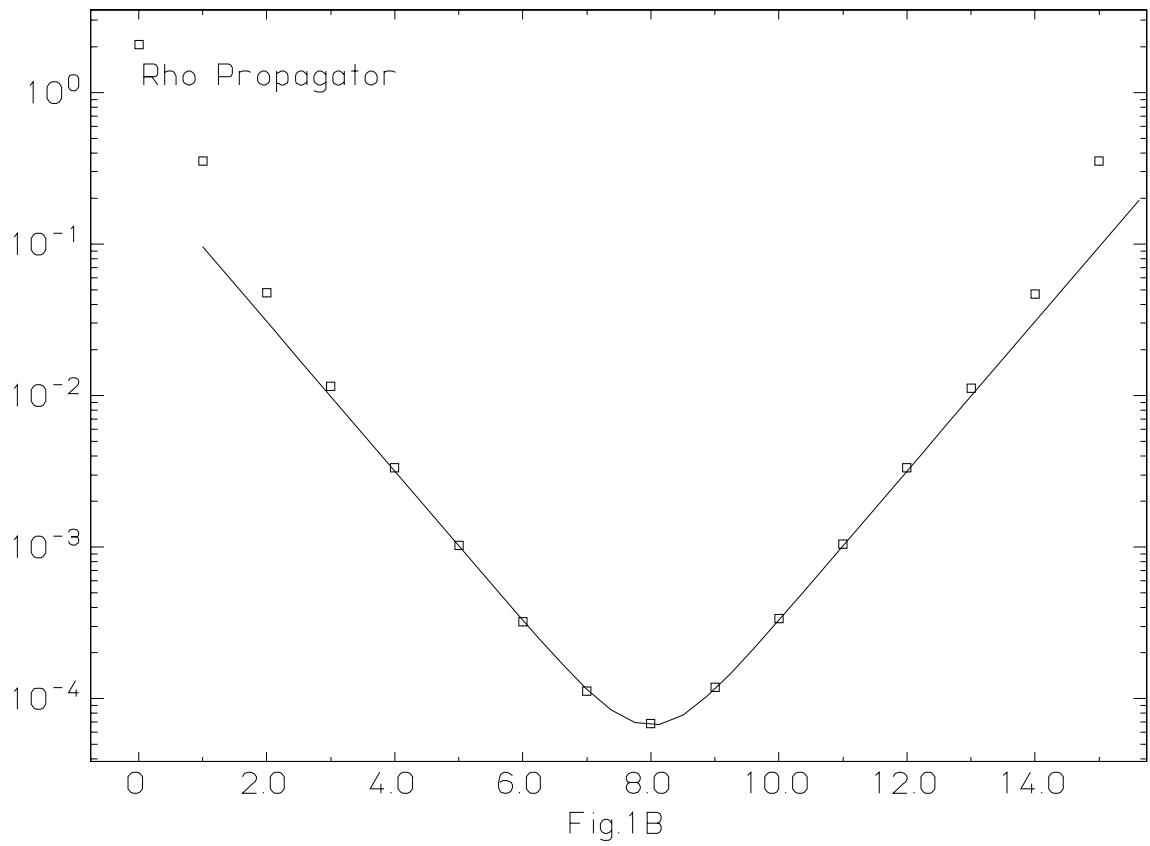
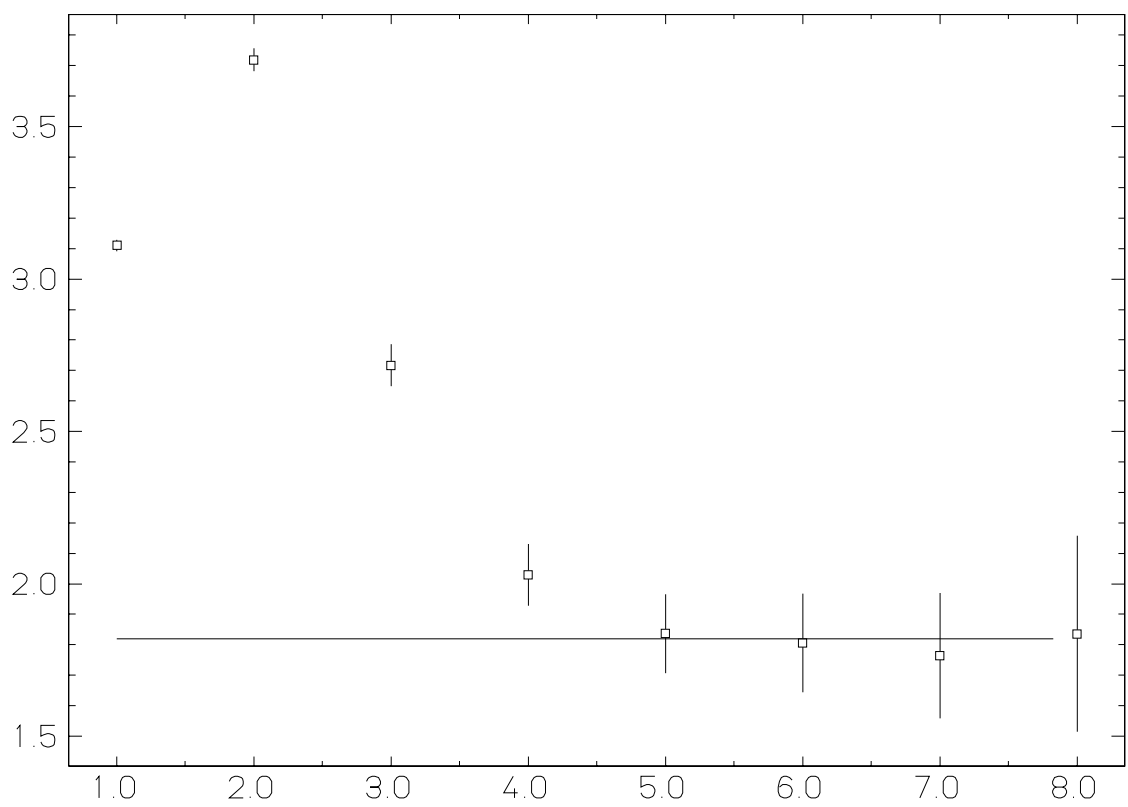
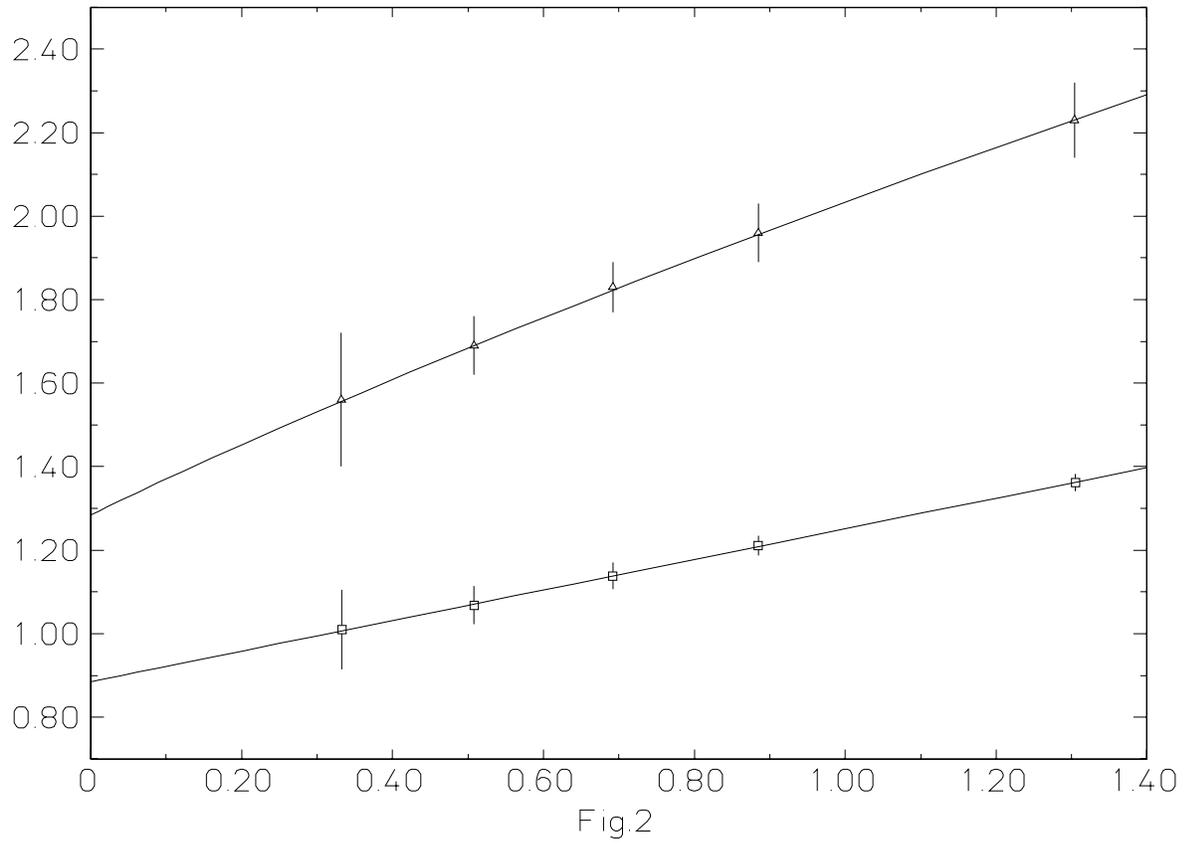
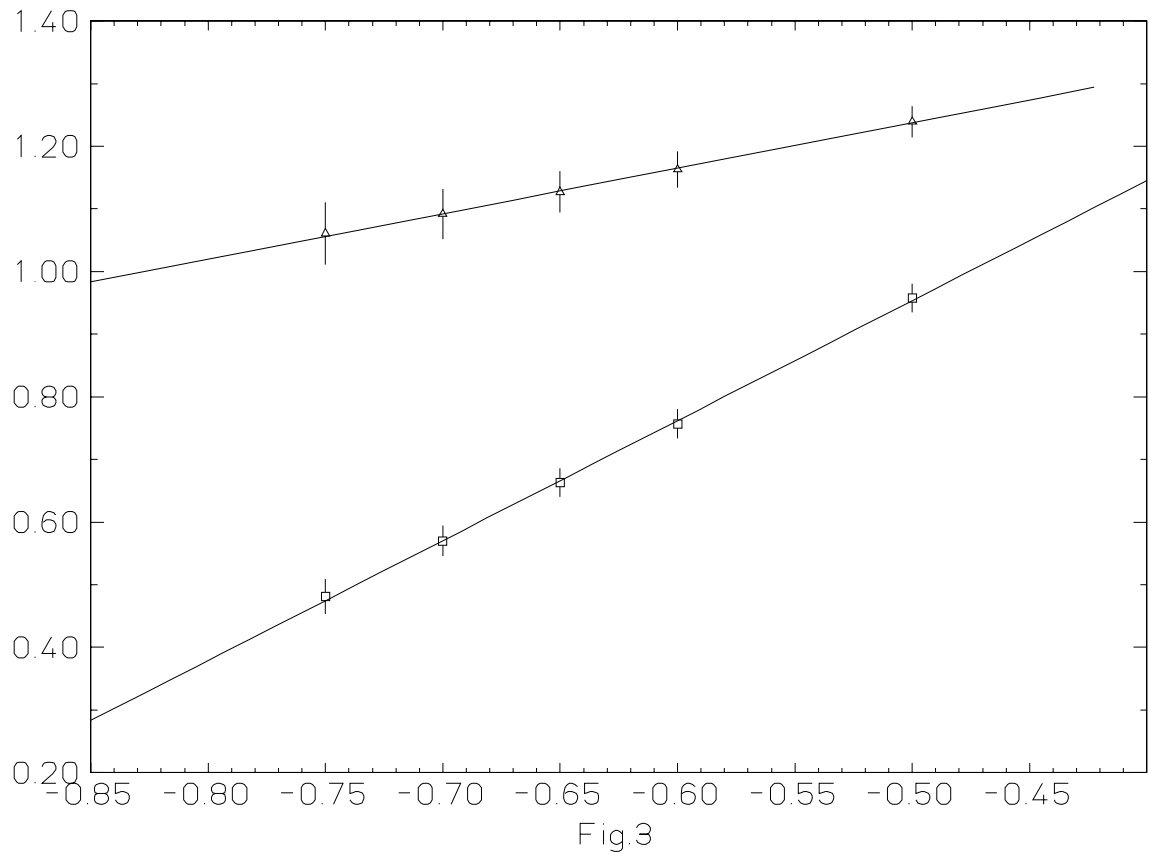
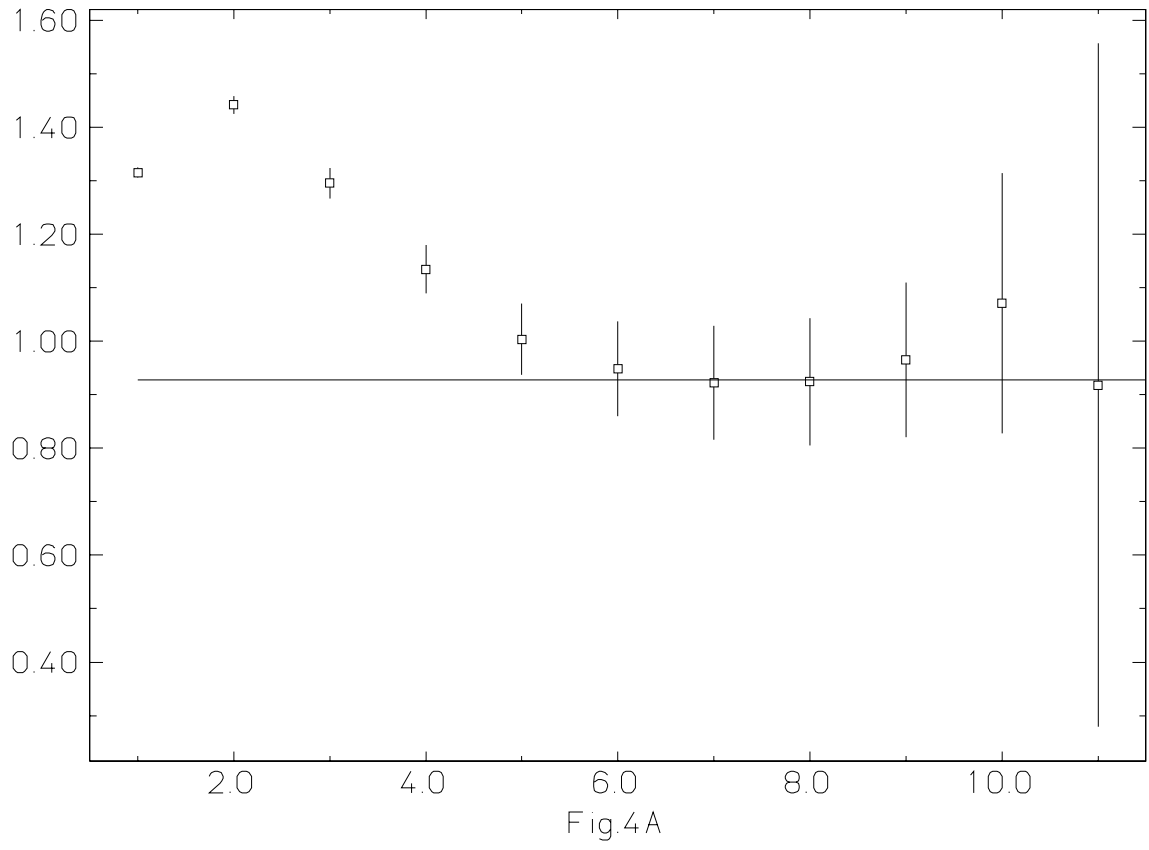
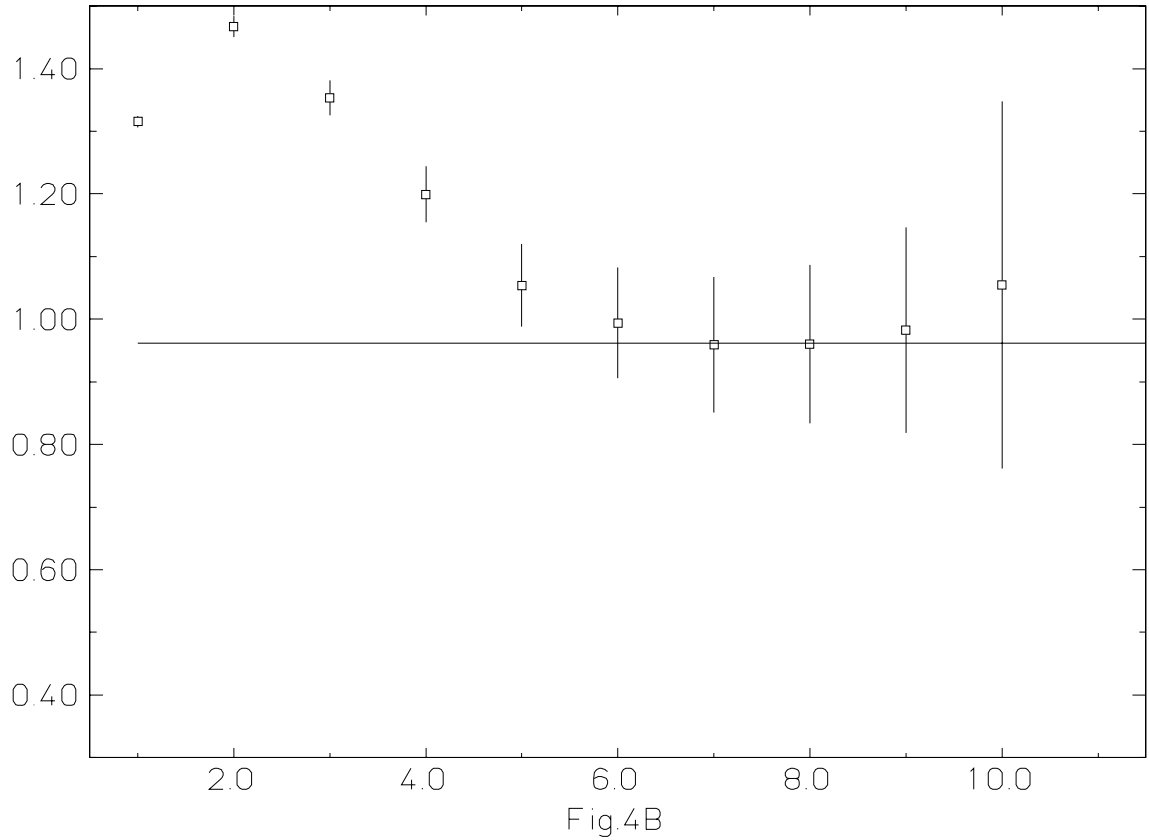
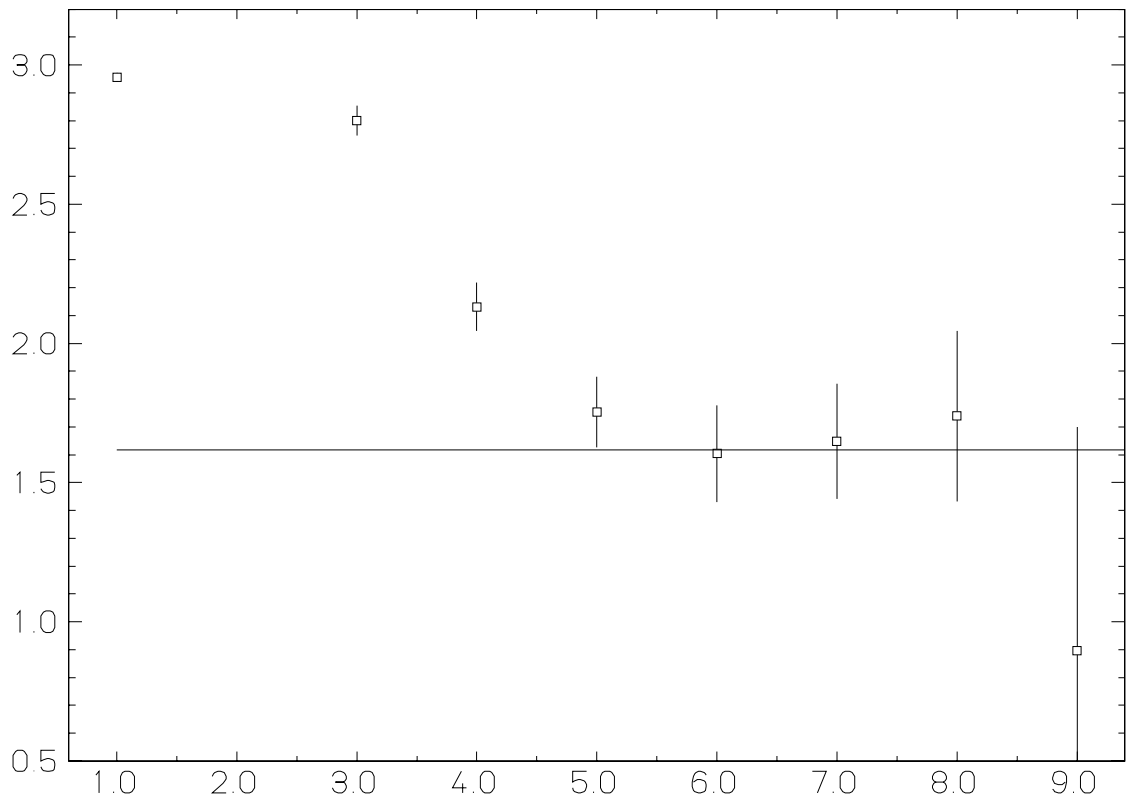
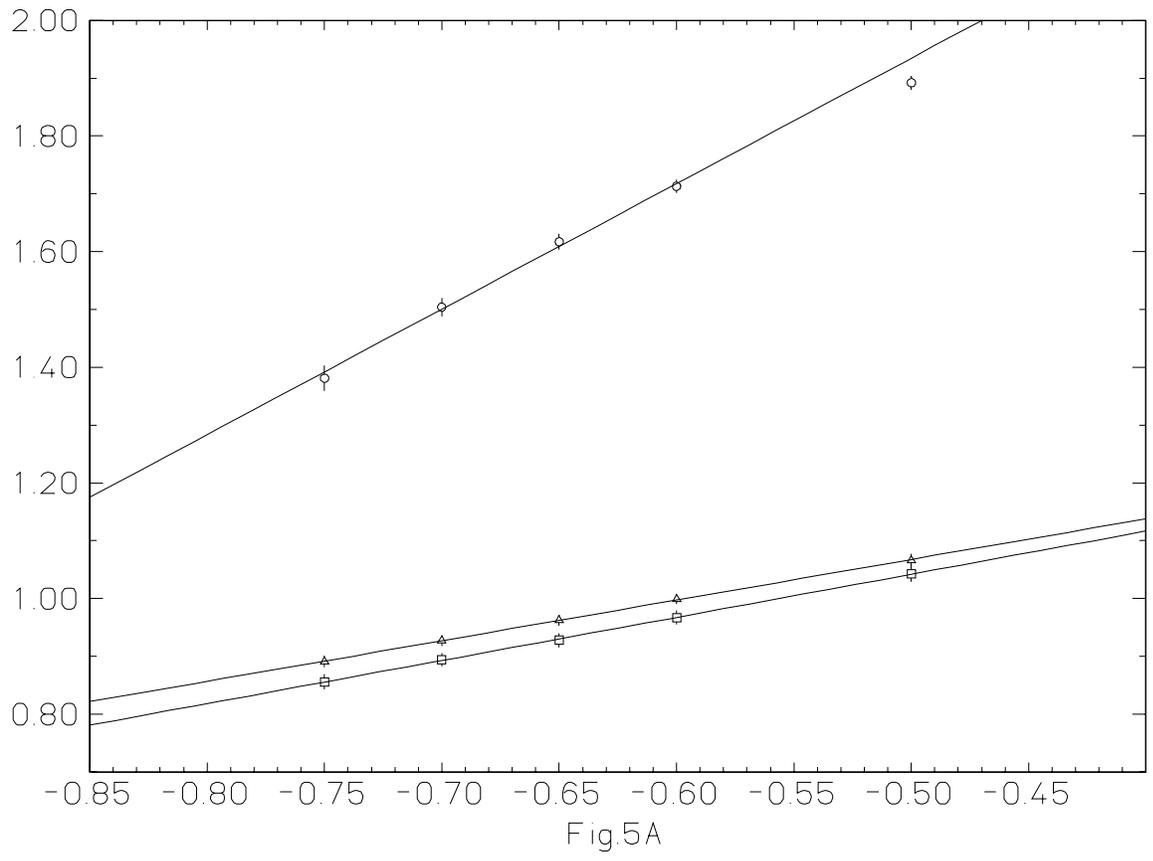
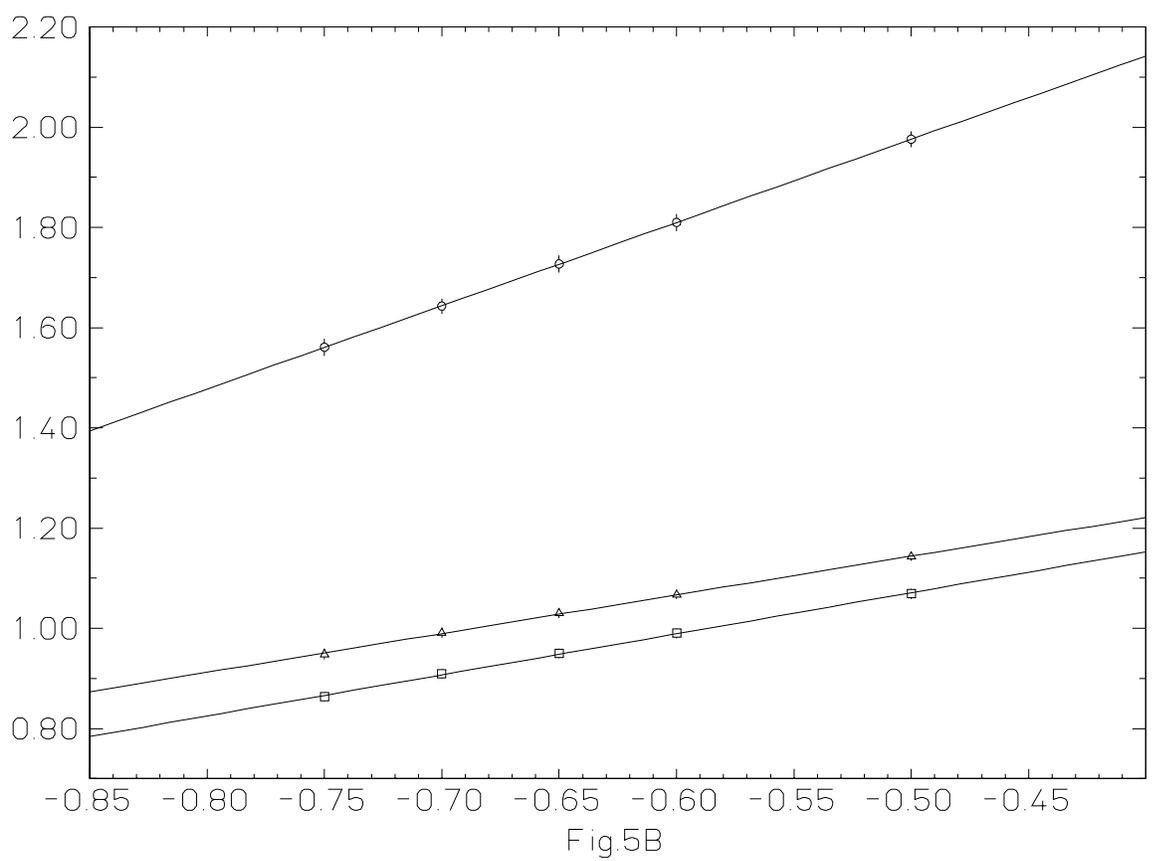
\end